\documentclass[twocolumn,   aps]{revtex4}

\usepackage{graphics}
\def \beq {\begin{equation}}
\def \eeq {\end{equation}}

\begin{document}

\title{Extracellular polymeric bacterial coverages as \\ minimal area surfaces}

\author{Alberto Saa}
\email{asaa@ime.unicamp.br}
\affiliation{Departamento de Matem\'atica Aplicada, IMECC - UNICAMP,  
13083-859 Campinas, SP, Brazil}

\author{Omar Teschke}
\email{oteschke@ifi.unicamp.br}
\affiliation{Laborat\'orio de NanoEstruturas e Interfaces, Instituto de F\'isica - UNICAMP, 
13083-970  Campinas, SP, Brazil}

\begin{abstract}
Surfaces formed by extracellular polymeric substances
enclosing individual and some small communities
of {\it Acidithiobacillus ferrooxidans} on plates of hydrophobic silicon and 
hydrophilic mica 
are analyzed by means of atomic force microscopy imaging. Accurate nanoscale 
descriptions of such coverage surfaces are 
obtained. The good agreement with the predictions of a rather simple but realistic 
theoretical model allows us to conclude that they correspond, indeed,  to
  minimal area (constant mean curvature) surfaces enclosing a given volume   
associated with the encased bacteria. This is, to the best of our knowledge, the first
shape characterization   of the coverage formed by these biomolecules, with potential
applications to the study of biofilms.
\end{abstract}

\maketitle

Extracellular polymeric substances (EPS) are produced by microorganisms during the process of 
adhesion to an environmental surface, acting mainly to protect 
them and to facilitate their interactions \cite{Evans}. The exact functions of EPS have not been 
completely elucidated yet because of their extremely heterogeneous nature. 
It is known, however, that EPS play significant roles in the formation and function of 
microbial aggregates, including matrix structure formation and microbial physiological 
processes\cite{EPS}. In this Note, we report an analysis, based experimentally on  
 atomic force microscopy (AFM) imaging, of the EPS bacterial 
 coverage produced in communities 
of {\it Acidithiobacillus ferrooxidans}  adhered to flat plates 
of silicon (hydrophobic) and mica (hydrophilic).  AFM  has the 
ability to image the coverage 
surface morphology in aqueous conditions, without any chemical fixation. 
  In particular, AFM has recently proved to be useful in imaging the morphology of 
bacteria\cite{r2}, liposomes\cite{3a}, and DNA molecules\cite{4a} on solid surfaces. 
Beech {\em et al.}\cite{Beech}, furthermore,  
showed that AFM allows the estimation of the width and height of bacterial exopolymeric 
capsule and bacterial flagella. We notice also that AMF has been recently used to
characterize wetting morphologies on microstructured surfaces\cite{Seemann}.
 
As we will show, AFM can be also used to determine the shape of different 
EPS coverage patterns of individual bacterium and some small communities of 
{\it A. ferrooxidans}. 
The appearance of the minimal area phenomena on extracellular polymeric coverage  
is associated with the need of the bacteria to   prevent losing of
  water   under drying conditions. The EPS secreted in solution or after 
fixation will have to cover the bacteria if they are going to survive. 
Since EPS production costs resources and 
energy to the bacteria, it would be natural to expect that EPS coverage 
surfaces should obey some 
variational principle, implying, therefore, that the observed surfaces should be {\em minimal} 
with respect to some criteria. At this scale ($\sim$1 $\mu$m), on the other hand, 
one does not expect any other force to be relevant besides of surface tension\cite{11a,12a,CR}. 
Consequently, the observed surfaces should correspond to minimal area 
(constant mean curvature) surfaces enclosing a 
given and fixed volume, associated, of course, with the encased bacteria. In this way, the 
observed surfaces would minimize both the potential elastic energy and the total amount of EPS 
necessary to form them. Our analyses confirm this hypothesis, EPS
coverages of {\it A. ferrooxidans} adhered to mica and silicon plates can be indeed
understood as minimal area surfaces enclosing some fixed volumes. For a review of the 
biological significance of {\em free}  ({\em i.e.}, without any
volume constraint, zero mean curvature)
minimal surfaces, see \cite{12a,CR}. Our conclusions are in 
agreement with the recently reported studies\cite{Seemann} on the minimal area
surfaces associated to different wetting morphologies on microstructured surfaces.
\begin{figure}[h]
\resizebox{0.65\linewidth}{!}{\includegraphics*{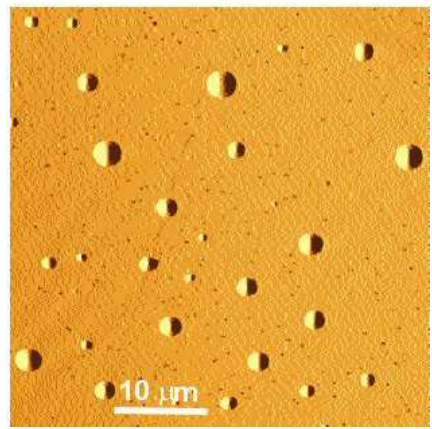}} 
\caption{{\it A. ferrooxidans} adhered to a hydrophilic mica plate observed in air.
For all bacteria, the covering material form  a cap-like structure. } 
\label{fig1}
\end{figure}

EPS form a highly 
entangled hydrated structure, composed basically by sugars and water linked by 
hydrogen bonds, in   agreement with our optical observations suggesting
that the EPS coverage behaves as a isotropic gel-like structure. 
Also, the typical EPS coverage has a 
volume 20 times larger than the encased bacterial volume.
The details about the bacteria growing conditions, EPS chemical composition analysis, and
the experimental setup are given in the Section Material and Methods of the
Supplementary Material.
 
\begin{figure*}[ht]
 \resizebox{0.3\linewidth}{!}{\includegraphics*{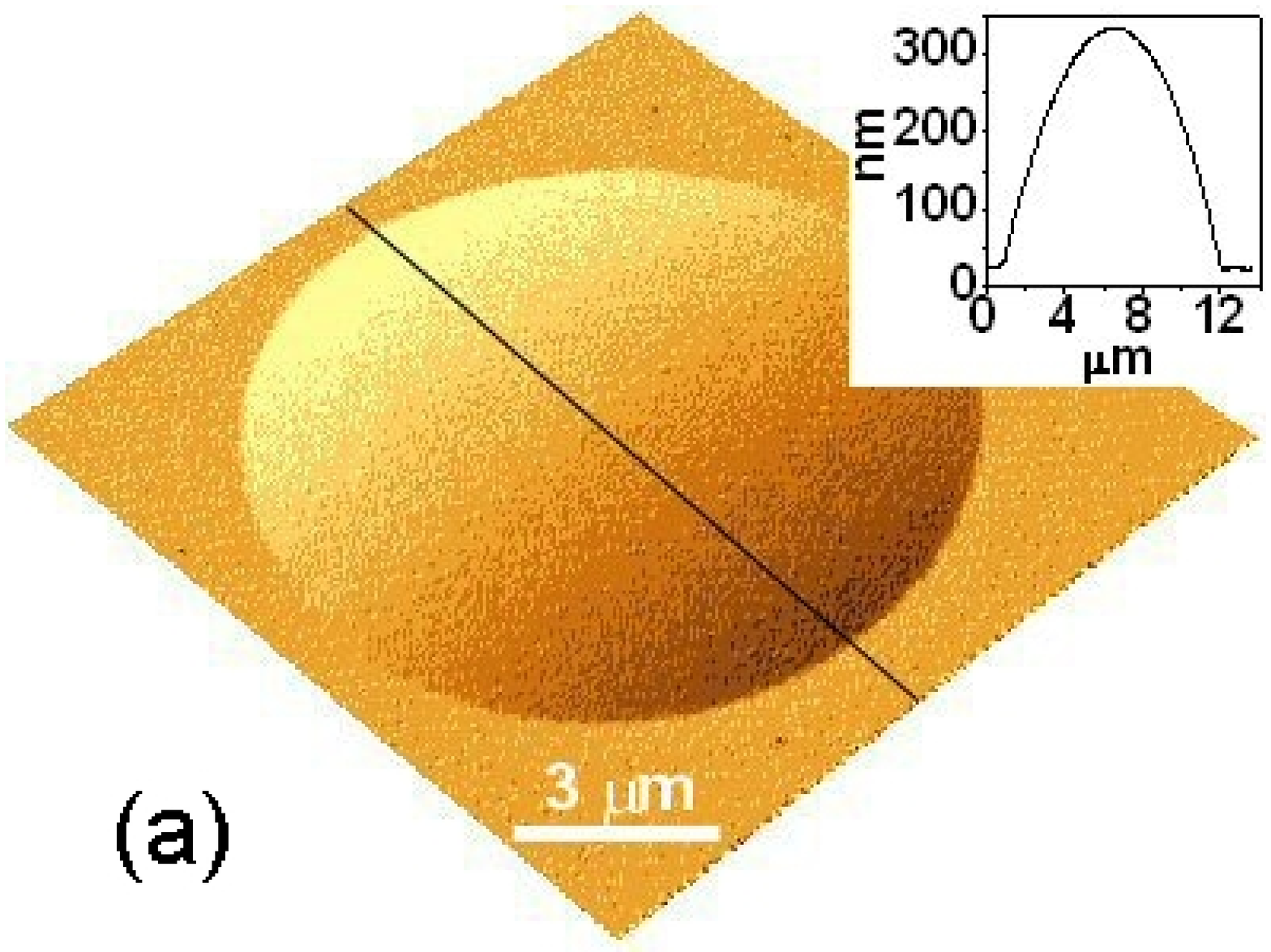}} 
 \resizebox{0.3\linewidth}{!}{\includegraphics*{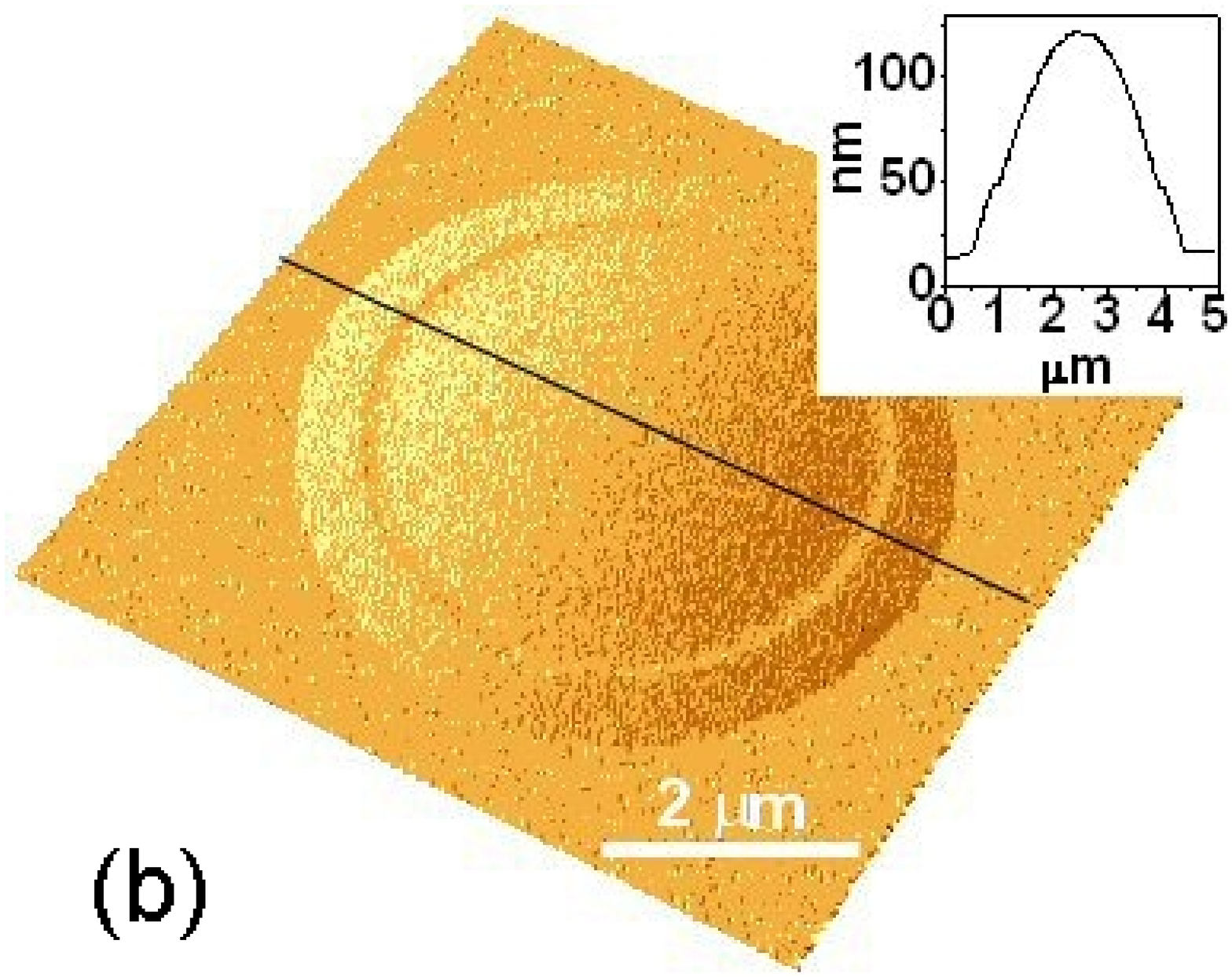}}
 \resizebox{0.3\linewidth}{!}{\includegraphics*{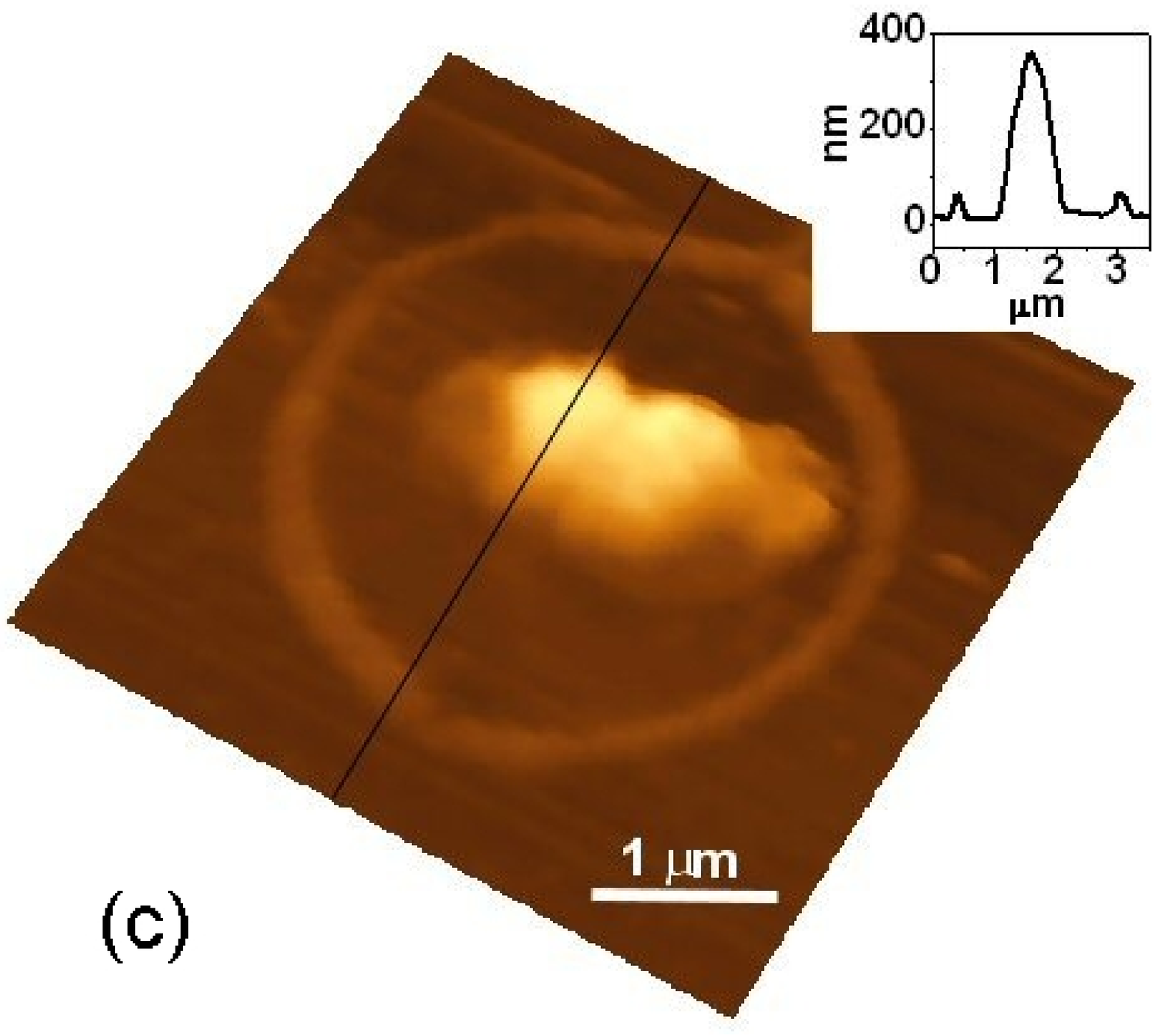}}
 \resizebox{0.3\linewidth}{!}{\includegraphics*{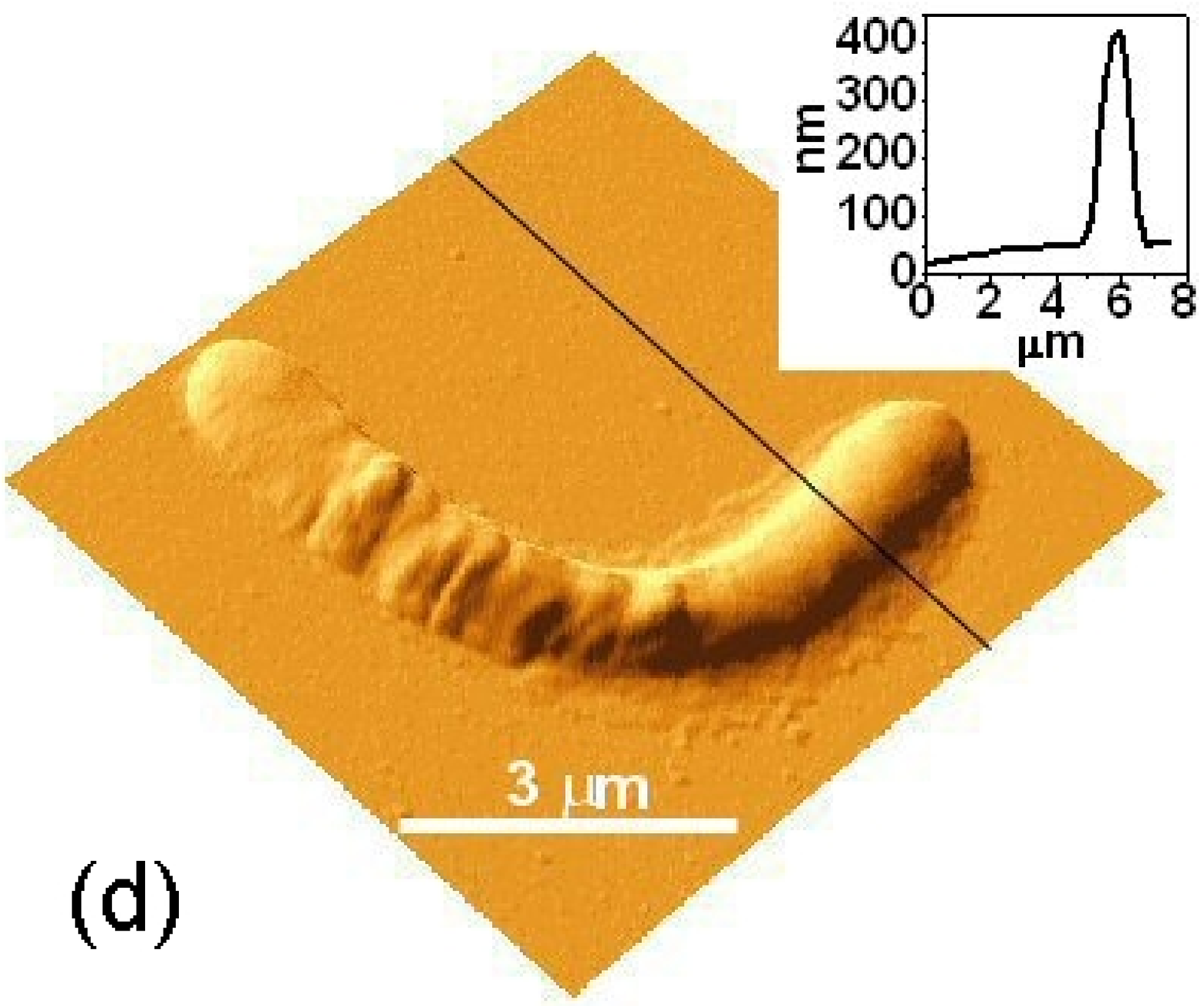}}
 \resizebox{0.3\linewidth}{!}{\includegraphics*{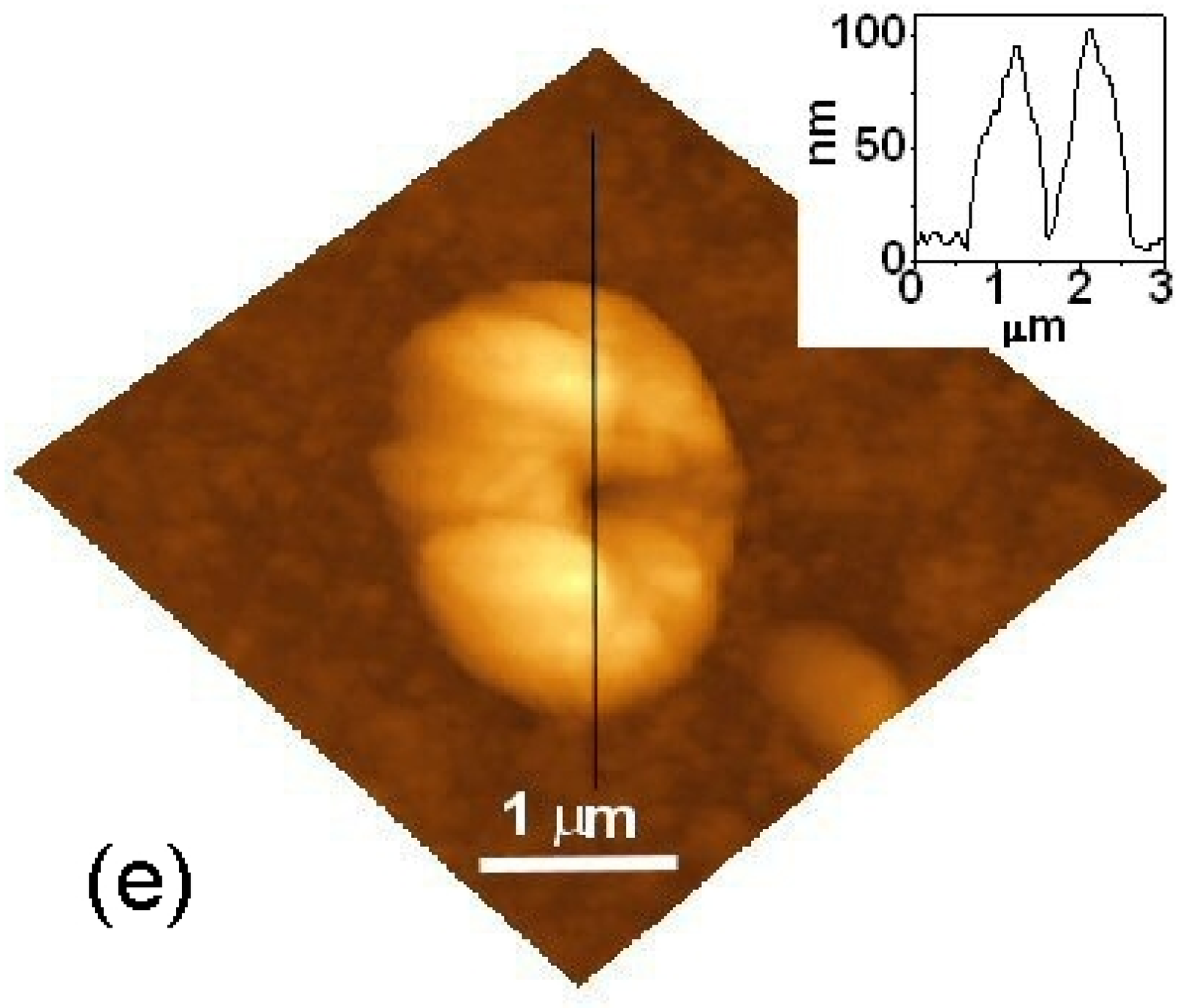}}
 \resizebox{0.3\linewidth}{!}{\includegraphics*{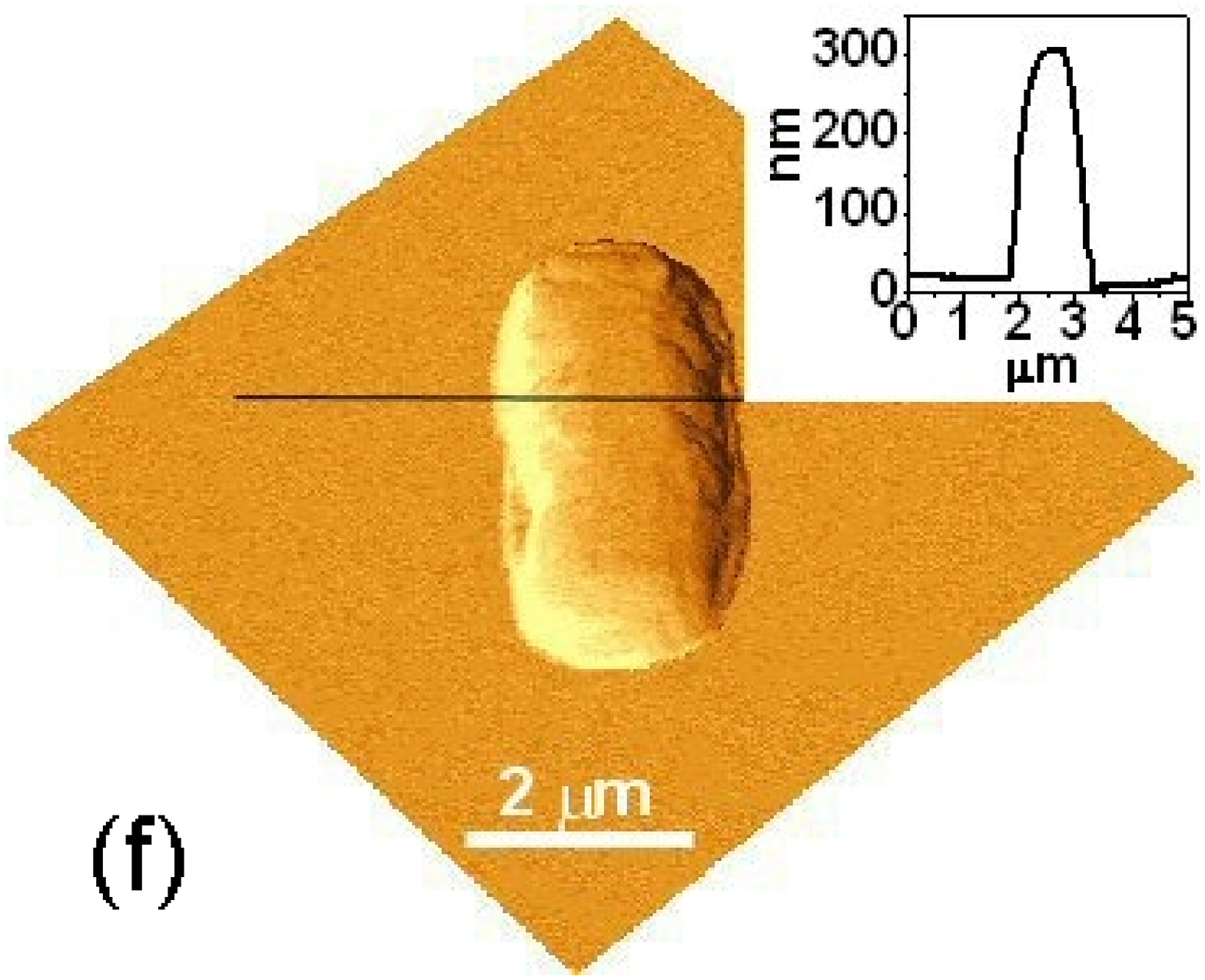}}
\caption{Some AFM images of the EPS coverage of single {\it A. ferrooxidans} adhered to hydrophilic
mica (a-c) and hydrophobic silicon (d-f) plates. 
The curves correspond  to the slices indicated in the images. All images except  (c) were obtained in aqueous
condition. Fig.  (c) corresponds to a rinsed sample similar to  (b), where one can
see clearly the circular ring corresponding to the thicker basis of the coverage. 
See the text for further details.}
\label{fig2}
\end{figure*}

Fig. \ref{fig1} shows a typical distribution of bacteria adhering to a 
hydrophilic mica plate at the 
center of the deposited droplet ($\sim 5$ mL, 3.0 pH),   corresponding to a view on 
an area of $\sim 2500$ $\mu$m$^2$. We notice that, typically,
each bacterium is isolated from 
the others and the area is almost uniformly covered. The shapes of the covering structures are
shown in detail  in Fig. 2, where   top views of   individual covered bacteria are 
displayed; all bacteria show a cap-like structure formed by the covering material. 
Also, images recorded after scanning large areas provide direct evidence for the presence 
of a continuous layer covering part of the substrate. Although layers as thin as $\sim 20$ nm are 
observed, most of the covered substrate has thicker layers ($\sim 600$ nm) of deposited 
material. We remind that 
bacteria shape is determined by their membranes structure, which typically 
have highly fluid shapes, varying from   cylindrical when  in a 
solution to some flat prolate structures when deposited on a substrate, as 
shown, for instance, in the dried sample depicted in   Fig. 2c.

The problem  of finding a minimal area surface enclosing a given 
volume is a classical isoperimetric (isovolume) variational problem, and several mathematical 
and computational tools are available to solve it in the most generic contexts.
Further details of our mathematical analysis can be found in the Section 
Minimal Area Surfaces of the Supplementary Material.
For the case of axisymmetric surfaces, analytical solutions are available, 
whereas for the non-symmetric case we had to use some approximations or 
iterative numerical methods. 
For the first case, 
one has yet two qualitative distinct cases according to the nature of the support region  
$\cal D$. For simply connected $\cal D$, it is well known that the minimal surface enclosing
a given volume corresponds to a spherical cap. Some surfaces with multiply connected support
can be well approximated by  a segment of a torus.
As we will see, such minimal area surfaces with 
multiply connected support will be useful to enlighten some of the observed structures, 
notably for the case of hydrophobic silicon plates.

Fig. \ref{fig2} shows some observed images of the EPS coverage for single {\it A. ferrooxidans}. 
The typical surface image for the mica 
plates is axisymmetric and has simply connected support. 
Table I in the Supplementary Material shows the relevant parameters, including the
observed and predicted values of the contact angle $\theta$, for several 
samples in mica plates with nearly circular support $\cal D$. 
Another property of the spherical cap that can be 
properly checked is that its intersection with any plane must be circle arcs. 
Fig. 1 in the Supplementary Material  shows the $\chi^2$ fitting of all the samples considered in
Table I.

An interesting image of an axisymmetric EPS coverage surface, 
often observed for {\em A. ferrooxidans} 
on mica plates, is that one showed in Fig. 2(b). It consists of a spherical cap born by a 
thicker circular ring, which, incidentally,
also has its external surface well described by a spherical 
segment. After rinsing these samples, one can see clearly (Fig. 2(c)) that the ring is composed by 
a less soluble EPS. The ring shape is very suggestive of a structure similar 
to the minimal toric segment (see the Supplementary Material). We notice also that such a kind 
of ring structure seems to be generally formed during   drying processes, 
possibly induced by capillary flows, see \cite{coffee}.

For the case of hydrophobic silicon plates, the images   are typically more irregular when
compared with those ones observed for   hydrophilic mica.
This can be understood recalling that, while for a hydrophilic substrate  the EPS produced
by the adhered bacteria can spread over easily, occupying large areas if compared
with the bacteria size, 
for a hydrophobic substrate the
produced EPS do tend to clump and to be very sensitive to eventual
surface microdefects of the substrate. Nevertheless, 
some of these non-symmetric coverages  can be understood with the help of the
axisymmetric minimal area surfaces with multiply connected support. It is the case, for instance,
of the image presented in Fig. 2(d). 
It corresponds to a segment of a long and curved   
figure. Sections as that one indicated in Fig. 2(d) are nearly circle arcs, 
but some samples have shown up with a high degree of irregularity. The longitudinal 
curve is also well described by a circle arc. 
The question about why the coverage has such a shape 
has no easy answer. We foresee basically 
two possibilities. Such a disposition could correspond to some frustrate minimal toric 
segment, or maybe an initially symmetric situation, as, for instance, a section of a 
long prolate figure, could 
evolve toward the non-symmetric situation due the surface stresses induced by the
microdefects of the hydrophobic substrate.
The coverage surface depicted in Fig. 2(e), despite of being not exactly symmetric, is
also very suggestive of a minimal toric segment. 
Minimal area surfaces with multiply connected support show up for the hydrophobic 
substrate presumably because the EPS coverage is prevented to evolve towards
the globally minimal area surface (the spherical cap) due to the difficulties of spreading over the
hydrophobic substrate. The EPS coverage tends, in this way, to some 
locally minimal area surface as, for instance, the toric segment.

The non-axisymmetric image displayed in 2(f) is a typical example of a coverage surface
we had to analyze
numerically. There is no analytical solution for the general 
non-symmetric case. However,
a plenty of numerical and semi-analytical methods are available to attack this problem. 
These EPS coverage surfaces shall be understood as  minimal area
surfaces enclosing a given volume and with a given region of support $\cal D$. 
There is no hope in solving this 
problem analytically for generic $\cal  D$. K. Brakke's public domain software 
SURFACE EVOLVER\cite{evolver}, nevertheless, has proved to be a powerful tool to this kind 
of problem. It is based on an iterative algorithm capable of find minimal area surfaces, according 
to quite general criteria, subjected to a given set of constraints, and for virtually any 
support $\cal D$, simply connected or not, convex or not. We have used it with success 
for solving our non symmetric cases. However, in these cases, 
due to the numerical nature of the solutions, we  are basically
restricted to some qualitative analysis. We have proceeded as follows. First, we
find a parametric representation of the boundary $\partial \cal D$. Then, EVOLVER
is ran for various volumes, until the maximal hight of the minimal area surface
constrained to $\partial \cal D$ coincides
with the observed coverage high. Then, a specific slice of the sample is compared with the 
results of EVOLVER. No $\chi^2$ tests could be properly done in this case. However,
a good agreement is observed. We notice that, by approximating the coverage 2(f) by
a very long symmetric prolate surface, the central slice should correspond also to a circle 
arc, see Fig. 2 in the Supplementary Material.

Biofilms\cite{Evans} are composed primarily of cells and EPS. Presumably, with some typical distance between the cells, 
their EPS coverage could touch each other without clumping together in a large and common coverage, forming
a EPS mesh that will certainly conditionate the physical properties of the associated biofilm. With the cells  
disposed in a quasi-regular way, such EPS mesh should resemble the periodic free minimal   surfaces\cite{12a,CR}.
In the same way that some important physical and biological properties of lipid-water phases, cell
membranes and biopolymers   are related to certain periodic free minimal   surfaces (see \cite{12a,CR} for further references),
one expects that relevant properties of a biofilm composed by such a EPS mesh might depend closely
on the geometrical details of the EPS coverage.
In this context, we notice that the complete mathematical solution of the 
problem of finding the minimal area surface  
enclosing and separating two given volumes (the double-bubble)
appeared only very recently\cite{2b}. These points are still
under investigation.

\acknowledgments

The authors are grateful to F.H.P. Knegt, R.F. Bergamo and L.M.M. Ottoboni for providing 
{\em A. ferrooxidans} bacteria cultures, 
M.E. Silva-Stenico and M.F. Fiore for the chemical quantification of carbohydrates, 
J. R. Castro and L. O. Bonugli for technical assistance,
B.M. Longo and J. Queiroz for valuable discussions, and the funding support of FAPESP (2003/12529-4) and 
CNPq (523.268/1995-5). 

\newpage

\begin{center}
{\Large\bf Supplementary Material} \\
\vspace{0.5cm}
\end{center}

\section{Material and Methods}

All the AFM images have been gotten by using an 
{\em A. ferrooxidans} strain LR\cite{13a} isolated 
from an acid effluent of the column leaching of uranium ore from Lagoa Real, BA, Brazil. 
The EPS composition has been analyzed
according to the phenol-sulfuric acid method using glucose standard\cite{Dubois}.
The bacterial suspension was deposited on a substrate and the total amount of
sugar was determined. It was shown that the amount of sugar
increased after a period of 24 hours. On the other hand, bacteria deposited
in a wet (water saturated) environment  did  not show any presence of sugar, 
suggesting strongly that sugars are the main component of EPS produced by the 
bacteria when adhered to a substrate. 
Moreover, EPS produced by bacteria hold several times their 
weight in water\cite{Roberson}. EPS form, hence, 
  a highly 
entangled hydrated structure, composed basically by sugars and water linked by 
hydrogen bonds, in perfect agreement with our optical observations suggesting
that the EPS coverage behaves as a isotropic gel-like structure. 
We observed also that the typical EPS coverage has a 
volume 20 times larger than the encased bacterial volume.

The observation of samples by AFM was always preceded by a visual observation   using 
the optical microscope attached to the AFM unit. Various areas are selected and then scanned with a 
large range of $\sim 50\times 50\ \mu{\rm m}^2$. When  regions with isolated bacterium are identified,
the 
scanning range is 
lowered down until only 1 bacterium is caught. Typically 50 bacteria 
are observed in a large scan range image. From this, a few are selected, usually those ones
with the best 
resolution.
Considering that 15 different sample preparations were used in this work,   that 5 samples were 
prepared from each run, and that 50 bacteria at least in each substrate were observed,  
  the typical   bacterial coverage is well represented by our images.

\subsection{Growth conditions and sample preparation}

The bacterial strain  was maintained  in modified 
TK liquid medium\cite{14a}: K$_2$HPO$_4$3H$_2$O, 0.4 g/L; MgSO$_4$.7H$_2$O, 0.4 d/L; 
(NH$_4$)$_2$SO$_4$, 0.4 g/L; FeSO$_4$7H$_2$O, 33.4 g/L; where the ferrous
sulfate has been included as an energy source\cite{Berg}.
The medium pH was 
adjusted to 1.8 by addition of sulfuric acid.
For culture growth, an inoculum of 
bacteria (5\% v/v) was added to 250 mL Erlenmeyers containing 100 mL of the modified 
TK medium. Growth was performed at $30^\circ$C under constant shaking at 150 rpm.

In order to study the topography of planktonic EPS, aliquots of liquid cultures 
containing 10 $\mu$L of a cell suspension that was grown to exponential phase were 
directly applied onto substrata. For the AFM imaging, a sample of 5 $\mu$L of cell 
suspension (approximately $10^9$ cells$/{\rm mL}$) was added to silicon or muscovite 
mica and air-dried for 2 h at $20^\circ$C  in an atmosphere with 60\% humidity.

\subsection{AFM images and techniques}

The model ThermoMicroscope AutoProbe CP\cite{CP}  was chosen for these experiments. 
The tip selected had a very small radius of curvature ($\sim 5$ nm), and the ultra-low spring 
constant of its cantilever ($\sim 0.03$ N/m) allowed probing live cells to suit our objectives 
without damage to the biological material. Before scanning, the size and shape of the AFM probe 
were characterized using a titanium reference sample from   ThermoMicroscope silicon grating\cite{CP}. 
The probe's cantilever was made of silicon nitride in a triangular shape 
of 200 $\mu$m in length.  Contact mode topographic images were recorded,   the imaging 
force was kept below 10 nN and the scan rate in the range of 1-4 Hz. 
There is no difference in the image for forward 
or reverse scanning. The two images may be recorded simultaneously but, since there no difference between 
then, usually only the forward scan is registered. Initially, we consider 
 views of $\sim 50\times 50\  \mu{\rm m}^2$ area,  with the bacteria 
  isolated from each other and randomly distributed over the substratum 
  ({\em i.e.} without large void regions). 
  A larger amplification depicts clearly coverages 
with a cap-like structure which can also show  a ticker ring surrounding the cap; the vertical profile image 
shows that this structure contains a single encased bacterial cell which is shown after washing the 
structure with water. A single cell has the dimensions of approximately 1 $\mu$m, whereas the 
typical diameter of 
the   EPS coverage is $4\ \mu {\rm m}$, indicating   that the ring is large 
enough to encircle 
a single bacterial cell. The images also show  that most of the structure that covers bacteria cell was 
removed when   washed with water, whereas  the rings of putative   insoluble material could not be 
removed.

The most common artifact in AFM image acquisition is the tip shape influence in the size of the images. 
However, the 5 nm radius tips used in this work have a negligible effect in determining the bacteria 
and EPS coverage   profiles, which have   typical size of a few micrometers. This situation was
studied in 
  \cite{calc}, where we consider  distortions and some other geometrical resolution limitations due to
  the response of   conical tips of different geometries.
Here, a  particularly more  serious drawback during AFM imaging was 
the tip pollution by EPS absorption. This problem was
partially solved   by using the above mentioned ultra-low spring
  cantilever. By adjusting the imaging force below 
 10 nN, samples could be scanned without the immersion of the tip end in the EPS layer. When a 
volume of EPS is attached to the tip, the result is a loss of image resolution, usually image widening or
even,
in some extreme cases,   the fixation of the tip to the substrate and the end of the scanning action. If the 
tip is dragging loose pieces of material, the results is the formation of parallel lines along the scanning 
direction. In any of the above cases, the scanning was interrupted and the tip   changed.

\section{Minimal Area Surfaces}

The problem\cite{elsgolts} of finding a minimal area surface enclosing a given 
volume is a classical isoperimetric (isovolume) variational problem, and several mathematical 
and computational tools are available to solve it in the most generic contexts.
The details of our mathematical analysis will be reported elsewhere\cite{OS2}.
The coverage surfaces are mathematically modeled by a smooth function 
$f(x,y)\ge 0$ with support restricted to a region $\cal D$, simply connected or not,
of the plane $(x,y)$. 
For the case of axisymmetric surfaces, analytical solutions are available, 
whereas for the non-symmetric case we had to use some approximations or 
iterative numerical methods. 
For the first case, by introducing appropriate polar coordinates $(\rho,\theta)$,
one has yet two qualitative distinct cases according to the nature of the support region  
$\cal D$. For simply connected $\cal D$, the minimal surface enclosing
a given volume corresponds to a spherical cap with equation
\beq
\label{arc}
(f(\rho) - d)^2 + \rho^2 = r^2,
\eeq
where $d$ and $r$ are free parameters.
 The volume of the associated spherical cap is given 
by $V_0 =  \frac{2\pi}{3}\left(  r^3 - d^3\right)  + \pi d  (r^2 -d^2)$,
and $\rho_{\rm max}^2 =  {r^2-d^2}$.
Hence, $\rho_{\rm max}$ and $V_0$ would be enough to determine unambiguous the surface. However, 
it is not a easy task to infer the volume enclosed by the surface from our AFM images, it 
is   more convenient, instead, to use   maximal 
high of the   surface $f(0)=r+d$  or the contact angle $\theta$ between the plate and the surface at 
the boundary $\partial\cal D$. These quantities  obey
\beq
\label{relat}
 \frac{2}{\tan\theta} = \frac{\rho_{\rm max}}{f(0)} -\frac{f(0)}{\rho_{\rm max}} .
\eeq
The relation (\ref{relat}) can be verified (and it could be falsified) easily from our 
axisymmetric images with simply connected $\cal D$. 
The spherical cap is the global minimum of the problem, 
{\em i.e.}, if no other constraint is imposed, the minimal area surface enclosing a given 
volume is the spherical cap.

The second axisymmetric case corresponds to the surfaces with multiply connected support.
The minimal area surface in this case is given by 
\beq
  \label{e18}
f(\rho) -d= \int_{\rho_{\rm min}}^\rho \frac{\frac{\lambda}{2}s^2 - c}
{\sqrt{s^2-\left(\frac{\lambda}{2}s^2 - c \right)^2}} \, ds,
\eeq
with $\rho_{\rm min} = (\sqrt{1+2\lambda c} -1)/\lambda$, where $d$,
$\lambda$ and $c$ are free   parameters.
Eq. (\ref{e18})
 can be expressed in a closed, but rather cumbersome, form by means of elliptic 
functions. A  
very interesting case occurs for $c \gg 1/\lambda$. In such a case, 
by introducing the new variable $s=\tau+\sqrt{2c/\lambda}$, $\tau \in [-1/\lambda,1/\lambda]$, 
Eq. (\ref{e18}) can be accurately approximated   by
\beq
\label{toric}
f( \sqrt{2c/\lambda} + \tau) -d\approx \int_{-\frac{1}{\lambda}}^\tau 
\frac{t}{\sqrt{\frac{1}{\lambda^2}-t^2}} \, dt = \sqrt{\frac{1}{\lambda^2}-\tau^2}
\eeq
Eq. (\ref{toric}) is easily recognized as the equation for a segment of a torus
with radii $\sqrt{2c/\lambda}$ and $1/\lambda$. 
As it has been mentioned, such minimal area surfaces with 
multiply connected support are useful to enlighten some of the observed structures, 
notably for the case of hydrophobic silicon plates.

\begin{table}[h]
\begin{tabular}{|r|r|r|r|r|r|r|r|r|r|}
\hline
Sample & $\rho_{\rm max}$ &  $f(0)$     &     ${\theta}_{\rm calc} $   &  $\theta_{\rm obs}$ & $-d\ \ $ &  $r\ \ \ $     & $\chi^2\ \ \ $ \\
       & $ {}^{(\mu m)} $ & ${}^{(nm)}$ &     ${}^{({\rm rad})} $    &   ${}^{({\rm rad})} $ &       $ {}^{(\mu m)} $ & $ {}^{(\mu m)} $ &  ${}^{(10^{-4})}$        \\
\hline
a$\;\,$ (43)  &3.497 & 193  &0.110 & 0.09 & 30.699 & 31.015 & 2.1 \\
b (158) &5.158&262&0.107&0.10& 51.452 & 51.868 & 3.5 \\
c (118)  &2.345&207&0.171&0.15 & 12.836 & 13.203 & 6.0 \\
d (124)  &1.379&67&0.097&0.09 & 14.206 & 14.322 & 1.0 \\
\hline
\end{tabular}
\caption{The relevant parameters, including the 
observed and calculated  contact angles  for some samples 
of EPS axisymmetric coverage on a mica plate. The values of $d$, $r$, and $\chi^2$ corresponds
to the  $\chi^2$ fitting of the spherical cap (Eqs. (\ref{arc}) and (\ref{relat})). Between parenthesis,
 in the first column,
we have the number of points read by AFM for each sample.}
\label{table}
\end{table}

\begin{figure}[ht]
\resizebox{ \linewidth}{!}{\includegraphics*{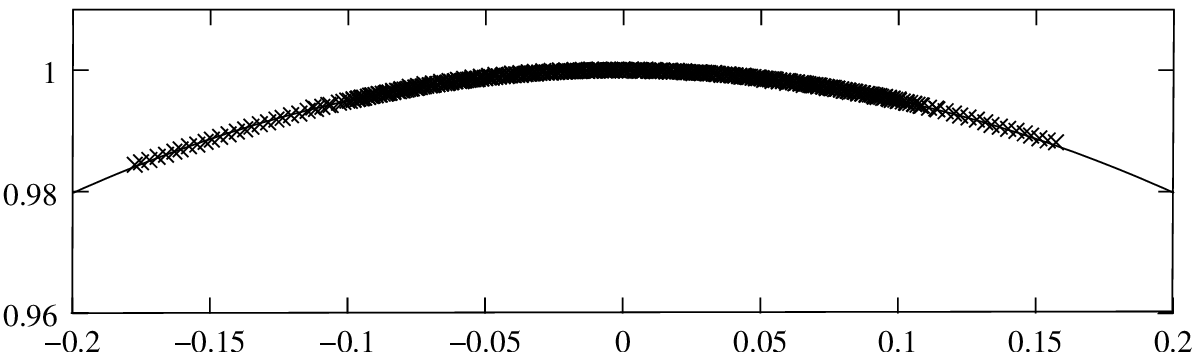}}
\caption{For illustrative purposes, all the 543 points of the samples considered in Table \ref{table} are
plotted on the unitary circle defined by   $x^2/r^2+(y-d)^2/r^2=1$ (the solid line), where
the values of $d$ and $r$ for each sample correspond to the  
$\chi^2$ fitting of  Table \ref{table}. The concordance is remarkable. }
\resizebox{ \linewidth}{!}{\includegraphics*{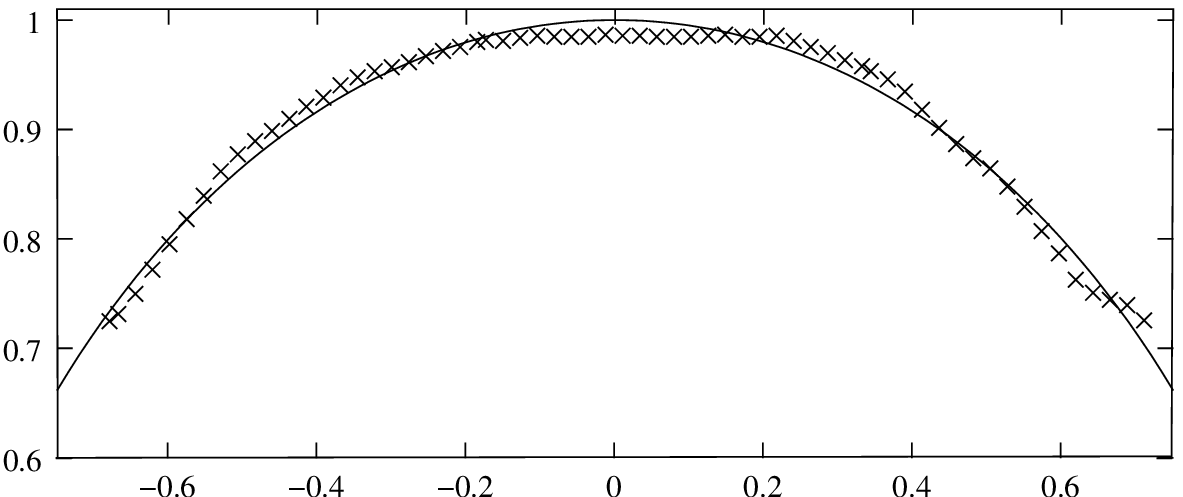}}
\caption{The $\chi^2$ fitting of the circle arc 
 $x^2/r^2+(y-d)^2/r^2=1$ (the solid line)
for the transversal slice
indicated in Fig. 2(f) of the text. 
The fitting corresponds to $r=1.043\, \mu$m and $-d=0.725\, \mu$m.
Circular transversal 
sections are good approximations for long symmetric
prolate figures. }
\end{figure}

\end{document}